# Effects of a Price limit Change on Market Stability at the Intraday Horizon in the Korean Stock Market

Wonse Kim*, Sungjae Jun†

**Abstract**

This paper investigates the effects of a price limit change on the volatility of the Korean stock market's (KRX) intraday stock price process. Based on the most recent transaction data from the KRX, which experienced a change in the price limit on June 15, 2015, we examine the change in *realized variance* after the price limit change to investigate the overall effects of the change on the intraday market volatility. We then analyze the effects in more detail by applying the discrete Fourier transform (DFT) to the data set. We find evidence that the market becomes more volatile in the intraday horizon because of the increase in the amplitudes of the low-frequency components of the price processes after the price limit change. Therefore, liquidity providers are in a worse situation than they were prior to the change.

___________________________________

* Graduate School, Department of Mathematics, Seoul National University, Seoul, Korea. E-mail: aquinasws@snu.ac.kr, Tel.: +82-10-7113-7851.

† Corresponding author: Courant Institute, New York University, New York, NY 10012. Email: sungjae.jun@nyu.edu, Tel: +82-16-223-4219.

**Introduction**

Emerging markets frequently adopt the price limit system, along with circuit breakers and trading halts, as a market-stabilization mechanism.[1] However, despite the recognized importance of the price limit system to policy, few theoretical trials have been conducted to explain the rationale of the price limit system in the equity market.[2] In addition, there is no agreement about the effect of the price limit system in academic research.[3] Given the lack of consensus on the desired effect of price limit system, policy-makers have changed the price limit several times in an attempt to achieve an optimal price limit that would efficiently stabilize the market, and several case studies have examined the effects the changes have had on market stabilization. Employing data from the Korea stock market (KRX), Berkman and Lee (2002) showed that the widening of price limits increases market volatility. In contrast, Chen (1993) used data from the Taiwan stock market to show that the widening of price limits decreased stock volatility in two of the three regimes studied. Using a data set similar to that Chen (1993) used, Kim (2001) showed that tightening price limits does not moderate stock market volatility.

Most studies in this line of research have investigated the effects of price limit changes on market stabilization using horizons longer than daily and have not addressed intraday analysis because markets are not considered to be considerably destabilized in such a short time. However, as the *Flash Crash* on May 6, 2010, showed, the market can be substantially destabilized in as little as about thirty minutes, so analyses of the intraday-horizon should be important parts of the study of market stabilization.[4] Even so, while policy-makers have tried to stabilize markets by changing their price limits, there has been no intraday analysis about the effect of these price limit changes on market stabilization.

---

[1] Lim and Park (2010) noted that, twenty-three of forty-three important markets use price limit systems.
[2] Lim and Park (2010), and Deb et al. (2010) developed theoretical models that explain the role of price limit systems in the presence of price manipulators.
[3] For studies that support proponents of the price limit system, refer to Arak and Cook (1997), Bildik and Elekdag (2004), and Ma et al. (1989a; 1989b). For studies that support opponents of the price limit system, refer to Bildik and Gulay (2006), Chen (1998), Fama (1989), and Kim and Rhee (1997).
[4] For example, Kirilenko et al. (2017) and Menkveld, and Yueshen (2016) investigated intraday price processes around and during the *Flash Crash*.

With this paper we seek to fill this research gap by investigating the effects of price limit changes on market stabilization in the intraday horizon. To this end, we exploit the most recent intraday transaction data from the KRX, from July 1, 2014, to May 31, 2016. This dataset provides an opportunity to test the effect of price limit changes empirically on intraday volatility, as the KRX's expansion of its price limits on June 15, 2015, is in the middle of the period included in our dataset. Based on our dataset, we examine the change in *realized variance*, introduced in Andersen, et al. (2001), after the price limit change to investigate the overall effects of the change on the intraday market volatility. We then analyze the effects in more detail by applying the discrete Fourier transform (DFT) to the data set.

Using DFT in addition to estimating realized variance provides several advantages in investigating intraday volatility over using the realized variance alone: First, whereas traditional intraday volatility measures such as realized variance assume a specific stochastic process model as the price process,[5] DFT is fully model-free and nonparametric, so analyses based on DFT can give results that are more realistic than those from an analysis based on traditional volatility measures. In addition, as Malliavin and Mancino (2002) pointed out, DFT is more robust than the traditional methods because DFT is based on integrating time series, while the algorithms used in estimating the most traditional volatility measures are usually based on a differentiation procedure.[6] Therefore, our additional analysis based on DFT provides a more realistic, robust, and detailed view on intraday volatility than would traditional approaches that estimate specific intraday volatility measures. To the best of our knowledge, this study is the first to investigate empirically the effect of a price limit change on market stabilization in the intraday horizon.

---

[5] Andersen et al.'s (2001) *realized variance*, which is the most popular measure of intraday volatility, assumes that price processes follow geometric Brownian motion.

[6] According to Malliavin and Mancino (2002), most traditional volatility measures use an algorithm based on a "differentiation procedure" that is highly unstable.

1. Data

We used transaction data from the KRX from July 1, 2014, through May 31, 2016. This data contains basic information like price, quantity traded, and time. Since the price limits changed only once during our sample period, from ±15 percent to ±30 percent on June 15, 2015, our sample period covers about eleven months before the price limit change and about eleven months after it. Our study considers only the KRX's largest 200 companies based on their market capitalizations as valued on June 15, 2015, the day the price limits changed. On that day, our sample stocks accounted for about 80 percent of total market capitalizations, so we excluded small cap stocks from our sample data. A year earlier, on June 15, 2014, the average market capitalization of our sample stocks was about 5.9 trillion Korean Won (about 5 billion USD), and they were traded an average of about 11,400 times.

2. Methodology

To investigate the overall change in the intraday market volatility in response to the price limit change, we estimated *realized variance*, as introduced in Andersen et al. (2001), under geometric Brownian motion, the parametric assumption about stock price processes. To calculate the realized variance for stock *i* on day *t*, we take the log differences of prices observed at the end of each five-minute interval, using our transaction tick data, to calculate five-minute returns. The realized variance of stock *i* on day *t*, $\sigma_{i,t}^2$, then, is defined as the sum of the squared five-minute returns:

$$\sigma_{i,t}^2 = \sum_{k=1}^{70} r(i,t)_k^2 , \qquad (1)$$

where $r(i,t)_k^2$ is the squared *k*-th five-minute log return of stock *i* on day *t*. With a daily transaction record from 9:00 to 14:50, there are seventy five-minute returns for each day.

Based on the $\sigma_{i,t}^2$ for each stock *i*, we calculated the time series average realized volatility for the periods before and after the price limit change. The time series average realized volatility *before* the price limit change of stock *i*, $\bar{\sigma}_{i,Before}$, is

$$\bar{\sigma}_{i,Before} = \sqrt{\frac{\sum_{t \in S.P.B} \sigma_{i,t}^2}{|S.P.B|}}, \qquad (2)$$

where $S.P.B$ is a set of days in our sample period that belongs to the period before the limit change, and the notation $|S|$ refers to the number of elements in a set $S$. The time series average realized volatility *after* the limit change of stock $i$, $\bar{\sigma}_{i,After}$ is

$$\bar{\sigma}_{i,After} = \sqrt{\frac{\sum_{t \in S.P.A} \sigma_{i,t}^2}{|S.P.A|}}, \qquad (3)$$

where $S.P.A$ is a set of days in our sample period that belongs to the period after the price limit change. By calculating the change rate of the time series averages of the realized volatilities of stock $i$, $RV_i \left(= \log \frac{\bar{\sigma}_{i,After}}{\bar{\sigma}_{i,Before}}, \right)$, we can investigate the effect of the limit change on the intraday volatility for stock $i$.

Next, to analyze the effects of the price limit change on the intraday market volatility in more detail, we applied DFT to the intraday price time series. We collected prices every one minute to get a 349-element price vector[7] for each stock $i$ on day $t$, $P_{i,t}$. The $j$-th element of the vector, $P_{i,t}$, is the log of the price of stock $i$ at $j$ minutes after the market opens on day $t$. Then we calculated Fourier coefficients using equations (4) and (5) for stock $i$ on day $t$:

$$a_{i,t}(w) = \frac{2}{349} \sum_{j=1}^{349} P_{i,t}(j) \cdot cos(2\pi w j/349) \qquad (4)$$

$$b_{i,t}(w) = \frac{2}{349} \sum_{j=1}^{349} P_{i,t}(j) \cdot sin(2\pi w j/349) \qquad (5)$$

$for\ w = 1, 2, \cdots, 174,$

where $P_{i,t}(j)$ is the $j$-th element of the vector $P_{i,t}$. Hence, we obtained a 174-element amplitude vector of stock $i$ on day $t$, $C_{i,t}$. The $w$-th element of the vector $C_{i,t}$ is the amplitude of frequency $w$, such that $C_{i,t}(w) \ (= \sqrt{a_{i,t}(w)^2 + b_{i,t}(w)^2}\ )$.

---

[7] In our sample period, the regular KRX market is open for 350 minutes, from 9:00 a.m. to 2:50 p.m. We obtain a 349-dimension price vector array by removing two prices: the opening price at 9:00 a.m. and the price at 2:50 p.m.

Using the calculated $C_{i,t}$, for each stock $i$, we calculated the time series average amplitude of each frequency component for the periods before and after the price limit change. Specifically, the time series average amplitude of the $w$-frequency component for the period *before* the price limit change of stock $i$, $B_{i,w}$, is

$$\overline{C_i(w)}_{Before} = \sqrt{\frac{\sum_{t \in S.P.B} C_{i,t}(w)^2}{|S.P.B|}}\ , \qquad (6)$$

where $S.P.B$ is a set of days in our sample period that belongs to the period before the limit change, and $C_{i,t}(w)$ is the $w$-th element of the vector $C_{i,t}$. Similarly, the time series average amplitude of $w$-frequency component for the period *after* the limit change of stock $i$, $A_{i,w}$, is

$$\overline{C_i(w)}_{After} = \sqrt{\frac{\sum_{t \in S.P.A} C_{i,t}(w)^2}{|S.P.A|}}\ , \qquad (7)$$

where $S.P.A$ is a set of days in our sample period that belongs to the period after the price limit change.

By calculating the change rate of the time series averages of the $w$-frequency components of stock $i$, $F_{i,w}\left(= \log \frac{\overline{C_i(w)}_{After}}{\overline{C_i(w)}_{Before}}\right)$, we can investigate the effect of a price limit change on the $w$-frequency component for stock $i$.

## 3. Results

- Insert Table 1 about here. -

Table 1 reports the cross-sectional averages of the change rate of the realized volatility of stock $i$, $RV_i$, for our sample stocks with their t-statistics for three periods: from February 2, 2015, to October 30, 2015; from October 1, 2014, to February 29, 2016; and from July 1, 2014, to May 31, 2016. The change rates of the realized volatility show significantly positive averages of 3.6 percent, 9.3 percent, and 8.13 percent, respectively, at the 1% significance level. Therefore, the empirical results in Table 1 show that the KRX became more volatile in the intraday horizon after the price limit change.

- Insert Figure 1 about here. -

Figure 1 shows the cross-sectional average change rate of the *w*-frequency component of stock *i*, $F_{i,w}$, with error bars for each *w*-frequency ($w = 1,2…174$) for our sample stocks for three periods: from February 2, 2015, to October 30, 2015; from October 1, 2014, to February 29, 2016; and from July 1, 2014, to May 31, 2016. The error bars represent 2.3 standard errors of the cross-sectional averages,[8] which corresponds to a confidence interval of about 99%. The three error bar graphs show that only in the low frequencies do the change rates of stocks' frequency components have significantly positive averages at the 1% level. This result indicates that the price fluctuations that originated from around low-frequency components became significantly stronger after the price limit change than they were before it, while the price fluctuations that originated from around high-frequency components were not affected by the change in the price limit.

## 4. Conclusions

We used the realized volatility and the DFT to investigate the effects of a price limit change on intraday price volatility by exploiting data from the KRX, which experienced a price limit change on June 15, 2015. We present two central findings: First, the realized volatility increased after the price limit change, suggesting that the market became more volatile in the intraday horizon after the price limit change. Second, among the various components of price processes, only the low-frequency components increased after the price limit change. Therefore, the inventory risk of liquidity-providers like market makers and retailers, who provide liquidity in the market and take the risk of adverse selection, increases following price limit increases because they risk increased low-frequency price swings even with the same inventories. This result suggests that the KRX became less favorable to liquidity providers after the price change and that this reduced favorability did not enhance market stabilization in the intraday horizon.

---

[8] For each *w*-frequency ($w = 1,2…174$), the standard error is calculated as the standard deviation of $F_{i,w}$ divided by the square root of the number of sample stocks, 200.

Table 1. Cross-sectional averages of the change rate of the realized volatility of stock $i$, $RV_i$.

| 2015/02/02 ~ 2015/10/30 | 2014/10/01 ~ 2016/02/29 | 2014/01/01 ~ 2016/05/31 |
| --- | --- | --- |
| 0.0360 | 0.0930 | 0.0813 |
| (3.5792) | (9.2614) | (8.3045) |

Note: *t*-statistics are reported in parentheses.

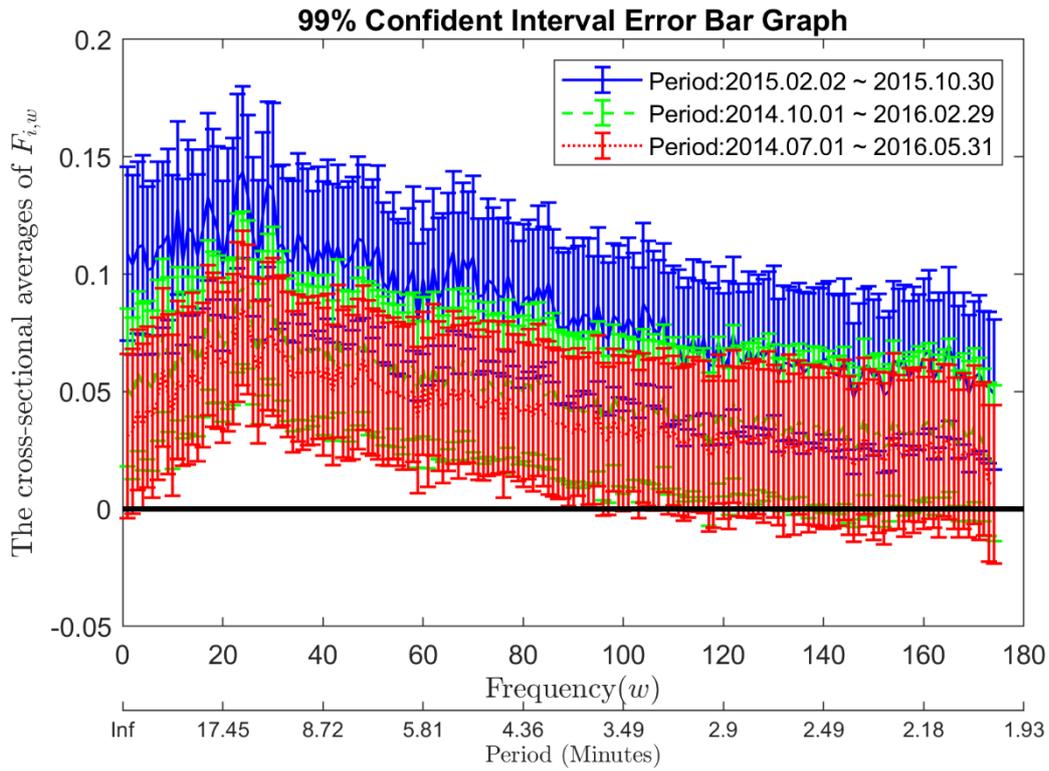

[Figure 1]

Fig. 1. Cross-sectional averages of $F_{i,w}$ for three time periods with 99% confidence intervals: from February 2, 2015, to October 30, 2015; from October 1, 2014, to February 29, 2016; and from July 1, 2014, to May 31, 2016.